\documentstyle[aps,prl,epsfig]{revtex}
\draft

\begin{document}
\markright{Garay, Anglin, Cirac, and Zoller.\hspace{3mm}
Sonic black holes in dilute Bose-Einstein condensates}
\title{Sonic black holes in dilute Bose-Einstein condensates}
\author{L.J.~Garay$^{1,2}$, J.R.~Anglin$^{1,3}$, J.I.~Cirac$^1$,
and P.~Zoller$^{1}$}
\address{$^1$ Institut f\"ur Theoretische Physik, Universit\"at
Innsbruck, Technikerstrasse 25, A-6020 Innsbruck, Austria}
\address{$^2$ Instituto de Matem\'{a}ticas y F\'{\i}sica Fundamental,
CSIC, C/ Serrano
121, E-28006 Madrid, Spain}
\address{$^3$ Institute for Theoretical Atomic and Molecular
Physics, Harvard-Smithsonian Center for Astrophysics, 60 Garden
Street, \\ Cambridge Massachussets 02135}
\date{30 May 2000}

\wideabs{
\maketitle

\begin{abstract}
The sonic analog of a gravitational black hole in dilute-gas
Bose-Einstein condensates is investigated. It is shown that
there exist both dynamically stable and unstable configurations
which, in the hydrodynamic limit, exhibit a behavior completely
analogous to that of gravitational black holes. The dynamical
instabilities involve creation of quasiparticle pairs in
positive and negative energy states. We illustrate these
features in two qualitatively different one-dimensional models,
namely, a long, thin condensate with an outcoupler laser beam
providing an ``atom sink,'' and a tight ring-shaped condensate.
 We have also
simulated the creation of a stable sonic black hole by solving
the Gross-Pitaevskii equation numerically for a condensate
subject to a trapping potential which is adiabatically deformed.
A sonic black hole could in this way be created
experimentally with state-of-the-art or planned technology.
\end{abstract}
\pacs{PACS number(s): 03.75.Fi, 04.70.Dy, 04.80.-y\hfill
{\it gr-qc/0005131; Phys. Rev. A 63, 023611 (2001)} }}

\section{Introduction}

Many investigations of dilute gas Bose-Einstein condensates are
directed towards experimentally creating nontrivial
configurations of the semiclassical mean field, or to predicting
the properties of such configurations in the presence of quantum
fluctuations. Such problems are hardly peculiar to condensates:
the quantum neighborhoods of interesting classical backgrounds
are important areas of research in most fields of physics. But
ultracold dilute gases are so easy to manipulate and control,
both experimentally \cite{BECexp95} and theoretically
\cite{dalfovo+99}, that they may
 allow us to decipher less amenable systems
by analogy. As an essay in such an application of condensates,
in this paper we discuss the theoretical framework and propose
an experiment to create the analog of a black hole in the
laboratory and simulate its radiative instabilities.

It is now commonly believed that, even in the context of
elementary particle physics, quantum field theory arises from a
still unknown underlying structure: it is an effective dynamical
theory, describing the low energy limit of collective phenomena
of the underlying microscopic theory. From this viewpoint, our
description of (more) fundamental phenomena, such as gravity or
electromagnetism, is actually similar to the theoretical
descriptions of many phenomena of condensed matter. To
understand superfluidity, superconductivity or dilute
Bose-Einstein condensation, we describe the dynamics of the
system in terms of collective modes (quasiparticles) whose
typical size is much larger than the distances between the
particles that constitute the underlying medium; but even
electrons and photons must be considered as the quasiparticles
of the deeper theory we do not yet know. In this sense we may
say that the major difference between our fundamental theories
and those we use in condensed matter is that in the latter case
the next microscopic level of description is actually well
understood.

With this fundamental background in mind, it is not so
surprising that condensed matter analogs of nontrivial
configurations appearing in relativistic quantum field theories
and gravitation can be constructed. For example, $^3$He has been
proposed as a laboratory counterpart of high-energy particle
physics. It has been argued that, under appropriate conditions,
excitations around the ground state of the system may resemble
the particle-spectrum of gauge theories of high energy physics
\cite{volovik99}. These condensed matter systems have also been
used to simulate topological defects characteristic of gauge
theories and which are considered to have played a cosmological
role in the early stages of the evolution of the universe such
as monopoles and cosmic strings \cite{volovik99}.

The past decade has witnessed an increasing interest in simulating
gravitational configurations and processes in condensed matter
systems in the laboratory. The key observation was originally made
by Unruh \cite{unruh81,unruh95} and further analyzed by Visser
\cite{visser,visser98}: phononic propagation in a fluid is
described by a wave equation which, under appropriate conditions,
can be interpreted as propagation in an effective relativistic
curved spacetime background, the spacetime metric being entirely
determined by the physical properties of the fluid under study,
namely, its density and flow velocity. Unruh urged a specific
motivation \cite{unruh81} for examining the hydrodynamic analogue
of an event horizon \cite{misner+73}, namely that as an
experimentally and theoretically accessible phenomenon it might
shed some light on the Hawking effect \cite{hawking74} (thermal
radiation from black holes, stationary insofar as the back
reaction is negligible). In particular, one would like to gain
insight into the role in the Hawking process of ultrahigh
frequencies~\cite{unruh95,jacobson91,corley+99}.

An event horizon for sound waves appears in principle wherever
there is a surface through which a fluid flows at the speed of
sound, the flow being subsonic on one side of the surface and
supersonic on the other. There is a close analogy between sound
propagation on a background hydrodynamic flow, and field
propagation in a curved spacetime; and although hydrodynamics is
only a long-wavelength effective theory for physical
(super)fluids, so also field theory in curved spacetime is to be
considered a long-wavelength approximation to quantum gravity
\cite{unruh95,visser98}. Determining whether and how sonic black
holes radiate sound, in a full calculation beyond the
hydrodynamic approximation or in an actual experiment, can thus
offer some suggestions about black hole radiance and its
sensitivity to high frequency physics (beyond the Planck scale).
The possibility that such high frequencies might have
consequences for observably low frequency phenomena is one of
the main reasons that black holes have deserved much attention:
there is reason to expect that an event horizon can act as a
microscope, giving us a view into physics on scales below the
Planck length. This is because modes coming from an event
horizon are redshifted into the low-energy regime as they
propagate out to be observed far away from the black hole.
Conversely, if we imagine tracking the observed signal back
towards its source, the closer we get to the horizon, the
shorter the wave length of the signal must be, until at the very
horizon we must either reach infinite energy scales, or
encounter a breakdown in general relativity and quantum field
theory in curved spacetime \cite{jacobson00}.

In understanding this problem, hydrodynamic and condensed matter
analogs of black holes may offer some of the experimental
guidance otherwise difficult to obtain in the case of
gravity\cite{jacobson00}. Under appropriate conditions and
approximations (which can be basically summarized in the
requirement the wave lengths of the perturbations be
sufficiently large), the propagation of collective fluctuations
(phonons) admits an effective general relativistic description,
in terms of a spacetime metric. This long wavelength regime
would correspond analogically to quantum field theory in curved
spacetime. The effective phonon metric may describe black holes,
as in general relativity, and so a phonon Hawking effect may be
possible; and certainly the problem of arbitrarily high
frequencies at the horizon is also present. But in this case,
when at short wave lengths the metric approximation is no longer
valid and a more microscopic theory must be used instead, the
accurate microscopic theory is actually known. And if the
hydrodynamic system is a dilute Bose-Einstein condensate, the
microscopic theory is actually tractable enough that we can make
reliable calculations from first principles. As we will argue,
trapped bosons at ultralow temperature can indeed provide an
analogue to a black-hole spacetime. Similar analogues have been
proposed in other contexts, such as superfluid helium
\cite{ruutu+96}, solid state physics \cite{reznik97}, and optics
\cite{leonhardt+00}; but the outstanding recent experimental
progress in cooling, manipulating and controlling atoms
\cite{cornell99} make Bose-Einstein condensates an especially
powerful tool for this kind of investigation.

The basic challenge of our proposal is to keep the trapped
Bose-Einstein gas sufficiently cold and well isolated to
maintain a locally supersonic flow long enough to observe its
intrinsic dynamics. Detecting thermal phonons radiating from the
horizons would obviously be a difficult additional problem,
since such radiation would be indistinguishable from many other
possible heating effects. This further difficulty does not arise
in our proposal, however, because the black-hole radiation we
predict is not quasistationary, but grows exponentially under
appropriate conditions. It should therefore be observable in the
next generation of atom traps, and may also raise new issues in
the theory of gravitational black holes.

In this paper, we extend and generalize the results of Ref.
\cite{bhbec}, including a more detailed analysis of the model
considered there as well as a new qualitatively different case.
The paper is organized as follows. In Sec. II, we show how sonic
horizons in dilute Bose-Einstein condensates may appear in the
hydrodynamic approximation, discuss the regime of validity of
such approximation, as well the validity of one-dimensional
models. Section III is devoted to the study of sonic horizons in
condensates subject to tight ring-shaped external potentials. We
present numerical results showing that both stable and unstable
black holes may be created under realistically attainable
conditions in current or near-future laboratories. We also study
the nature of the dynamical instabilities that appear for
certain configurations. In Sec. IV, we discuss a different
configuration, namely that of a sink-generated black hole in an
infinite one-dimensional condensate, and show that there also
exist black-hole configurations, although they are not stable.
We summarize and conclude in Sec. V. The Appendix is devoted to
the issues of redundancy and normalization of the dynamically
unstable Bogoliubov modes (associated with complex
eigenfrequencies).

\section{Sonic black holes in condensates}

A Bose-Einstein condensate is the ground state of a second
quantized many body Hamiltonian for $N$ interacting bosons
trapped by an external potential $V_{\rm ext}({\mathbf{x}})$
\cite{dalfovo+99}. At zero temperature, when the number of atoms
is large and the atomic interactions are sufficiently small,
almost all the atoms are in the same single-particle quantum
state $\Psi({\mathbf{x}},t)$, even if the system is slightly
perturbed. The evolution of $\Psi$ is then given by the
well-known Gross-Pitaevskii equation, which in appropriate units
can be written as
$$
i\hbar\partial_t\Psi=\left(-\frac{\hbar^2}{2m}\nabla^2+ V_{\rm
ext}+\frac{4\pi a\hbar^2}{m}|\Psi|^2\right)\Psi ,
$$
where $m$ is the mass of the individual atoms and $a$ is the
scattering length. The wave function of the condensate is
normalized to the total number of atoms $\int d^3{\mathbf{x}}
|\Psi({\mathbf{x}},t)|^2=N$.

Our purposes do not require solving the Gross-Pitaevskii
equation with some given external potential $V_{\rm
ext}({\mathbf{x}})$; our concern is the propagation of small
collective perturbations of the condensate, around a background
stationary state
$$
\Psi_s({\mathbf{x}},t)=\sqrt{\rho({\mathbf{x}})}
e^{i\vartheta({\mathbf{x}})} e^{-i\mu t/\hbar},
$$
where $\mu$ is the chemical potential. Thus it is only necessary
that it be possible, in any external potential that can be
generated, to create a condensate in this state. Indeed, many
realistic techniques for ``quantum state engineering,'' to
create designer potentials and bring condensates into specific
states, have been proposed, and even implemented successfully
\cite{cornell99}; and our simulations indicate that currently
known techniques should suffice to generate the condensate
states that we propose.

Perturbations about the stationary state
$\Psi_s({\mathbf{x}},t)$ obey the Bogoliubov system of two
coupled second order differential equations. Within the regime
of validity of the hydrodynamic (Thomas-Fermi) approximation
\cite{dalfovo+99}, these two equations for the density
perturbation $\varrho$ and the phase perturbation $\phi$ in
terms of the local speed of sound
$$
c({\mathbf{x}})\equiv\frac{\hbar}{m}\sqrt{4\pi
a\rho({\mathbf{x}})},
$$
and the background stationary velocity
$$
{\mathbf{v}}\equiv\frac{\hbar}{m}\nabla\vartheta
$$
read
$$
\dot\varrho=-\nabla\cdot\left(\frac{m}{4\pi a\hbar}c^2\nabla\phi
+{\mathbf{v}}\varrho\right)
\qquad
\dot\phi=-{\mathbf{v}}\cdot\nabla\phi -\frac{4\pi a\hbar}{m}\varrho.
$$
Furthermore, low frequency perturbations are essentially just
waves of (zero) sound. Indeed, the Bogoliubov equations
 may be reduced to a single second order equation for the
condensate phase perturbation $\phi$. This differential equation
has the form of a relativistic wave equation
$\partial_\mu(\sqrt{-g}g^{\mu\nu}\partial_\nu\phi)=0$, with
$g=\det g_{\mu\nu}$, in an effective curved spacetime with the
metric $g_{\mu\nu}$ being entirely determined by the local speed
of sound $c$ and the background stationary velocity
${\mathbf{v}}$. Up to a conformal factor, this effective metric
has the form
$$
(g_{\mu\nu})=\left(\begin{array}{cc} -(c^2- {\mathbf{v}}^2) &
-{\mathbf{v}}^{\rm T}\\ -{\mathbf{v}} & {\mathbf{1}}
\end{array}\right).
$$

This class of metrics can possess event horizons. For instance,
if an effective sink for atoms is generated at the center of a
spherical trap (such as by an atom laser out-coupling technique
\cite{andrews97}), and if the radial potential profile is
suitably arranged, we can produce densities $\rho(r)$ and flow
velocities ${\mathbf{v}}({\mathbf{x}})=-v(r){\mathbf{r}}/r$ such
that the quantity $c^2-{\mathbf{v}}^2$ vanishes at a radius
$r=r_h$, being negative inside and positive outside. The sphere
at radius $r_h$ is a sonic event horizon completely analogous to
those appearing in general relativistic black holes, in the
sense that sonic perturbations cannot propagate through this
surface in the outward direction
\cite{unruh81,unruh95,visser98}. This can be seen explicitly by
writing the equation for the radial null geodesics of the metric
$g_{\mu\nu}$:
$$
\dot r_{\pm}=-v\pm c,
$$
which can be obtained from setting the proper interval
$ds^2=g_{\mu\nu}dx^\mu dx^\nu$ equal to zero and restricting the allowed
motion to the radial direction, so that
$$
-(c^2-v^2) +2v\dot r +\dot r^2=0.
$$
The ingoing null geodesic $r_-(t)$ is not affected by the
presence of the horizon and can cross it in a finite coordinate
time $t$. The outgoing null geodesic $r_+(t)$ on the other hand
needs an infinite amount of time to leave the horizon since
$\dot r_+=0$ at the horizon. The physical mechanism of the sonic
black hole is quite simple: inside the horizon, the background
flow speed $v$ is larger than the local speed of sound $c$, and
so sound waves are inexorably dragged inwards.

In fact there are two conditions which must hold for this
dragged sound picture to be accurate. Wavelengths larger than
the black hole itself will of course not be dragged in, but
merely diffracted around it. And perturbations must have
wavelengths
$$
\lambda\gg\frac{\pi\hbar}{mc},\quad
\frac{\pi\hbar}{mc\sqrt{|1-v/c|}}.
$$
Otherwise they do not behave as sound waves since they lie
outside the regime of validity of the hydrodynamic
approximation. These short-wavelength modes must be described by
the full Bogoliubov equations, which allow signals to propagate
faster than the local sound speed, and thus permit escape from
sonic black holes. So, to identify a condensate state $\Psi_s$
as a sonic black hole, there must exist modes with wavelengths
larger than these lower limits (which in terms of the local
healing length $\xi({\mathbf{x}})\equiv\hbar/[mc({\mathbf{x}})]$
read $\lambda\gg 2\pi\xi,\ 2\pi\xi/\sqrt{|1-v/c|}$), but also
smaller than the black hole size. Even if such an intermediate
range does exist, the modes outside it may still affect the
stability of the black hole as discussed below.

As it stands, this description is incomplete. The condensate
flows continually inwards and therefore at $r=0$ there must be a
sink that takes atoms out of the condensate. Otherwise, the
continuity equation $\nabla (\rho{\mathbf{v}})=0$, which must
hold for stationary configurations will be violated. From the
physical point of view, such a sink can be accomplished by means
of an outcoupler laser beam at the origin. (Such outcouplers are
the basic mechanisms for making trapped condensates into ``atom
lasers,'' and they have already been demonstrated experimentally
by several groups. A tightly focused laser pulse changes the
internal state of the atoms at a particular point in the trap,
and can also be made to give them a large momentum impulse. This
ejects them so rapidly through the always dilute condensate
cloud that they do not significantly disturb it; effectively,
they simply disappear.)

\begin{figure}
\begin{center}
\epsfig{file=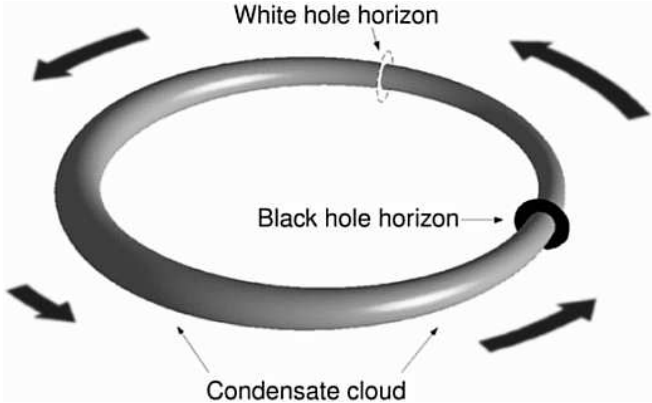,width=\columnwidth}
\end{center}
\caption{The tight ring-shaped configuration, with
both black and white horizons, and no singularity. Arrows
indicate condensate flow velocity, with longer arrows for faster
flow.}
\label{lopsided}
\end{figure}

\begin{figure}
\begin{center}
\epsfig{file=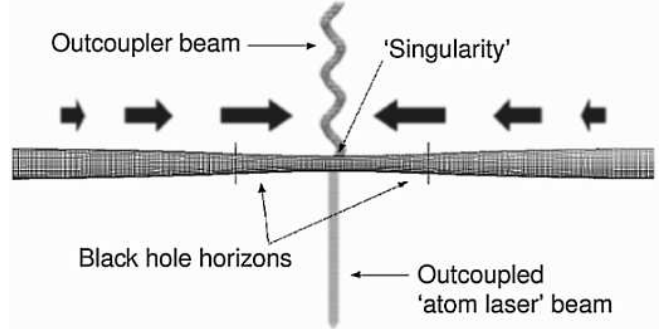,width=\columnwidth}
\end{center}
\caption{The tight cigar-shaped configuration, with two black hole
horizons and a `singularity' where condensate is outcoupled.
Arrows indicate condensate flow velocity, with longer arrows for
faster flow.} \label{leakypinch}
\end{figure}

We have analyzed several specific systems which may be suitable
theoretical models for future experiments, and have found that the
qualitative behavior is analogous in all of them. Black holes
which require atom sinks are both theoretically and experimentally
more involved, however; moreover, maintaining a steady transonic
flow into a sink may require either a very large condensate or
some means of replenishment. We will therefore first discuss an
alternative configuration which may be experimentally more
accessible and whose description is particularly simple: a
condensate in a very thin ring that effectively behaves as a
periodic one-dimensional system (Fig.~\ref{lopsided}). Under
conditions that we will discuss, the supersonic region in a ring
may be bounded by two horizons: a black hole horizon through which
phonons cannot exit, and a `white hole' horizon through which they
cannot enter. Then we will analyze another simple one-dimensional
model, of a long, straight condensate with an atom sink at the
center (Fig.~\ref{leakypinch}).

The existence of instabilities that do not show up in the
one-dimensional approximation is an important question in
condensate physics, which is under active theoretical and
experimental investigation. The essential principles have long
been clear, inasmuch as the current dilute condensates really are
the weakly interacting Bose gases that have been used as toy
models for superfluidity for several decades. The fact that actual
critical velocities in liquid helium are generally far below the
Landau critical velocity is understood to be due partly to the
roton feature of the helium dispersion relation, but this is not
present in the dilute condensates. Viscosity also arises due to
surface effects, however, and these may indeed afflict dilute
condensates as well. The point here is that in addition to the
bulk phonon modes considered by Landau, and quite adequately
represented in our one-dimensional analysis, there may in
principle be surface modes, with a different (and generally lower)
dispersion curve. If such modes exist and are unstable, it is very
often the case that, as they grow beyond the perturbative regime,
they turn into quantized vortices, which can cut through the
supercurrent and so lower it.

Whether or not such unstable surface modes actually exist in the
Bogoliubov spectrum of a dilute condensate is an issue that has
recently been analyzed both numerically and analytically, and it
is quite clear that such surface modes exist only if the
confining potential is quite rough (which is not only easy to
avoid with a magnetic or optical trapping field, but very hard
to achieve) \cite{winiecki}, or if the condensate dynamics in
the directions perpendicular to the flow is hydrodynamic. That
is, the condensate must be at least a few healing lengths thick,
so that surface modes decaying on the healing length scale can
satisfy all the required boundary conditions \cite{khawaja}. By
saying that we are considering an effectively one-dimensional
condensate, we mean precisely that this is not the case. For
instance, for the tight-ring model, in this regime, the radial
trap scale is the shortest length scale in the problem, and the
radial trap frequency is the highest frequency; this effectively
means that excitations of nontrivial radial modes, including
surface modes, are energetically frozen out. (In the limit of
radial confinement within the scattering length, our model
breaks down for other reasons --- but the scattering length can
easily be two orders of magnitude smaller than the healing
length). The issue of the supercurrent stability in tightly
confined ring shaped traps has been addressed in Ref.
\cite{svistunov} where the authors arrive at a positive
conclusion and also clarify the role of finite temperature and
possible trap anisotropy.

\section{Sonic black/white holes in a ring}

In a sufficiently tight ring-shaped external potential of radius $R$,
motion
in radial ($r$) and axial ($z$) cylindrical coordinates is
effectively frozen. We can then write the wave function as
$
\Psi(z,r,\theta,\tau)=f(z,r)\Phi(\theta,\tau)
$
and normalize $\Phi$ to the number of atoms in the condensate
$\int_0^{2\pi}d\theta|\Phi(\theta)|^2= N$, where with the
azimuthal coordinate $\theta$ we have introduced the
dimensionless time $\tau=(\hbar/mR^2)t$. The Gross-Pitaevskii
equation thus becomes effectively one-dimensional:
\begin{equation}
i\partial_\tau\Phi=\left(-\frac{1}{2}\partial_\theta^2+{\cal
V}_{\rm ext}+\frac{{\cal U}}{N} |\Phi|^2\right)\Phi,
\label{eq:gp-ring}
\end{equation}
where
$
{\cal U} \equiv 4\pi a N R^2 \int dzdrr |f(z,r)|^4
$
and ${\cal V}_{\rm ext}(\theta)$ is the dimensionless effective potential
(in which we have already included
the chemical potential) that results from the dimensional reduction.
The stationary solution can then be written as
$
\Phi_s(\theta,\tau)=\sqrt{\rho(\theta)} e^{i\int\!d\theta
v(\theta)}
$
and the local dimensionless angular speed of sound as
$c(\theta)=\sqrt{{\cal U}\rho(\theta)/N}$. Periodic boundary
conditions around the ring require the ``winding number''
$
w\equiv(1/2\pi)\int_0^{2\pi}d\theta v(\theta)
$
to be an integer.

The qualitative behavior of horizons in this system
is well represented by the two-parameter family of condensate
densities
$$
\rho(\theta)=\frac{N}{2\pi}(1+b\cos\theta),
$$
where $b\in[0,1]$. Continuity, $\partial_\theta(\rho v)=0$, then
determines the dimensionless flow-velocity field
$$
v(\theta)=\frac{{\cal U}w\sqrt{1-b^2}}{2\pi c(\theta)^2},
$$
which depends on $w$ as a third discrete independent parameter.
Requiring that $\Phi_s(\theta,\tau)$ be a stationary solution to
Gross-Pitaevskii equation then determines how the trapping
potential must be modulated as a function of $\theta$. All the
properties of the condensate, including whether and where it has
sonic horizons, and whether or not they are stable, are thus
functions of ${\cal U}$, $b$ and $w$. For instance, if we
require that the horizons be located at $\theta_h=\pm\pi/2$,
which imposes the relation ${\cal U}=2\pi w^2(1-b^2)$, then we
must have $c^2-v^2$ positive for $\theta\in (-\pi/2,\pi/2)$,
zero at $\theta_h=\pm\pi/2$, and negative otherwise, provided
that ${\cal U}<2\pi w^2$. The further requirement that
perturbations on wavelengths shorter than the inner and the
outer regions are indeed phononic implies ${\cal U}\gg2\pi$,
which in turn requires $w\gg1$ and $1\gg b\gg 1/w^2$. In fact,
detailed analysis shows that $w\gtrsim 5$ is sufficient.

\subsection{Stability}

The mere existence of a black hole solution does not necessarily
mean that it is physically realizable: it should also be stable
over sufficiently long time scales. Since stability must be
checked for perturbations on all wavelengths, the full
Bogoliubov \cite{dalfovo+99} spectrum must be determined. For
large black holes within large, slowly varying condensates, this Bogoliubov
problem may be solved using WKB methods that closely resemble
those used for solving relativistic field theories in true black
hole spacetimes \cite{corley+99}.  A detailed adaptation of
these methods to the Bogoliubov problem will be presented
elsewhere \cite{otherpaper}.  The results are
qualitatively similar to those we have found for black holes in
finite traps with low winding number, where we have resorted to
numerical methods
because, in these cases, WKB techniques may fail for just those
modes which threaten to be unstable.

Our numerical approach for our three-parameter family of
black/white holes in the ring-shaped condensate has been to
write the Bogoliubov equations in discrete Fourier space, and
then truncate the resulting infinite-dimensional eigenvalue
problem. Indeed, writing the wave funtion as $\Phi=\Phi_s
+\varphi e^{i\int d\theta v(\theta)}$,
decomposing the perturbation $\varphi$ in discrete modes
$$
\varphi(\theta,\tau)=\sum_{\omega,n} e^{-i\omega \tau} e^{in\theta}
A_{\omega,n}u_{\omega,n}(\theta)+ e^{i\omega^* \tau} e^{-in\theta}
A_{\omega,n}^*v_{\omega,n}^*(\theta),
$$
and substituting into the Gross-Pitaevskii equation we obtain
the following equation for the modes $u_{\omega,n}$ and
$v_{\omega,n}$:
$$
\omega \left(\begin{array}{c} u_{\omega,n} \\ v_{\omega,n}
\end{array}\right)=
\sum_p \left(\begin{array}{cc}
h^+_{np} & f_{np}\\ -f_{np} & h^-_{np}
\end{array}\right)
\left(\begin{array}{c} u_{\omega,p} \\ v_{\omega,p}
\end{array}\right).
$$
In this equation,
\begin{eqnarray}
f_{np}&=&\frac{1}{2\pi}\int_0^{2\pi} d\theta e^{-i(n-p)\theta}
 c(\theta)^2,
\nonumber\\
h_{np}^\pm&=&
\pm \frac{n^2}{2}\delta_{np}
+\frac{1}{2\pi}\int_0^{2\pi} d\theta e^{-i(n-p)\theta}
\nonumber\\
&&\times\left[ pv(\theta)-\frac{1}{2}v'(\theta)
\pm\left(c(\theta)^2+
\frac{1}{2}\frac{c''(\theta)}{c(\theta)}\right)\right],
\nonumber
\end{eqnarray}
which, after some lengthy calculations, can be written as
\begin{eqnarray}
f_{np}&=&\frac{{\cal U}}{2\pi}\left(\delta_{n,p}+ \frac{b}{2}
\delta_{n,p+1} + \frac{b}{2}
\delta_{n,p-1}\right),\nonumber\\
h^\pm_{np}&=&\frac{1}{2}(n+p)w\sqrt{1-b^2}\alpha_{n-p}
\nonumber\\
&\pm&\left(f_{np}+\frac{4n^2-1}{8}\delta_{n,p}+
\frac{1-b^2}{8}\beta_{n-p}
\right),
\nonumber
\end{eqnarray}
where
\begin{eqnarray}
\alpha_i&=&\sum_{j\geq |i|,\ i+j\ {\rm even}}^\infty
\left(\frac{-b}{2}\right)^j\left(\begin{array}{c} j\\
(i+j)/{2}\end{array}\right),\nonumber\\
\beta_i&=&\sum_{j\geq |i|,\ i+j\ {\rm even}}^\infty
\left(\frac{-b}{2}\right)^j\left(\begin{array}{c} j\\
(i+j)/{2} \end{array}\right) (j+1).\nonumber
\end{eqnarray}
 Eliminating Fourier components above a sufficiently high
cutoff $Q$ has negligible effect on possible instabilities,
which can be shown to occur at relatively long wavelengths. We
face then an eigenvalue problem for the $2(Q+1)\times 2(Q+1)$
matrix built out of blocks of the form
$$
\left(\begin{array}{cc}
h^+_{np} & f_{np}\\ -f_{np} & h^-_{np}
\end{array}\right).
$$
 The numerical solution to this eigenvalue equation,
together with the normalization condition
$\int d\theta (u^*_{\omega^*,n}u_{\omega',n'}- v^*_{\omega^*,n}v_{\omega',n'})
=\delta_{nn'}\delta_{\omega\omega'}$, provides the
allowed frequencies.
Real negative eigenfrequencies for modes of positive norm
are always present, which means that black hole configurations are
energetically unstable, as expected. This
feature is inherent in supersonic flow, since the speed of sound is
also the Landau critical velocity. In a sufficiently cold and
dilute condensate, however, the time scale for dissipation may
in principle be made very long, and so these energetic
instabilities need not be problematic \cite{shlyapnikov}.

\begin{figure}
\begin{center}
\epsfig{file=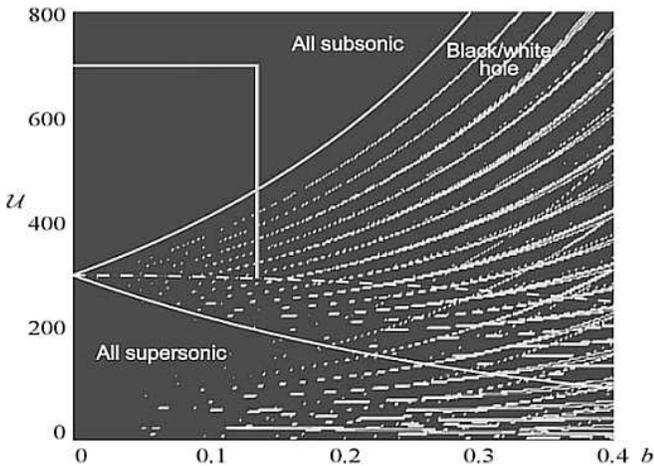,width=\columnwidth}
\end{center}
\caption{Stability diagram for winding number $w=7$. Solid
dark-grey areas represent the regions of stability. Smaller plots
at higher resolution confirm that the unstable `fingers' are
actually smooth and unbroken. Points on the dashed curve are
states with horizons at $\theta_h=\pm\pi/2$, so that the
black/white hole fills half the ring.} \label{fig:stab}
\end{figure}

\begin{figure}
\begin{center}
\epsfig{file=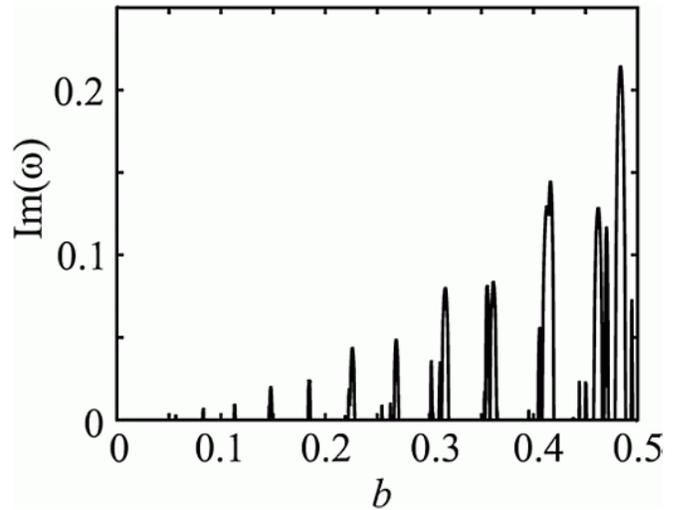,width=\columnwidth}
\end{center}
\caption{Stability diagram for black/white holes of maximum size,
i.e., along the dashed line of Fig. \ref{fig:stab}.}
\label{fig:stab2}
\end{figure}

More serious are dynamical instabilities, which occur for modes
with complex eigenfrequencies. Since the Bogoliubov theory is
based on a quantized Hamiltonian that is Hermitian, there are
certainly no complex energy eigenvalues; but the natural
frequencies of normal modes can indeed be complex [in which case
the usual rule, that energy eigenvalues are $\hbar(n+1/2)$ times
the mode frequencies, simply breaks down]. A detailed discussion
of the quantum mechanics of dynamical instability is presented in
the Appendix; for the purposes of our main discussion it suffices
to note that complex (mode) eigenfrequencies are indeed genuine
physical phenomena, and by no means a numerical artifact. For
sufficiently high values of the cutoff (e.g., $Q\geq 25$ in our
calculations), the complex eigenfrequencies obtained from the
 truncated eigenvalue problem
become independent of the cutoff within the numerical error. The
existence and rapidity of dynamical instabilities depend
sensitively on $({\cal U},b,w)$. For instance, see
Fig.~\ref{fig:stab} for a contour plot of the maximum of the
absolute values of the imaginary part of all eigenfrequencies
for $w=7$, showing that the regions of instability are long,
thin fingers in the $({\cal U},b)$ plane. Figure \ref{fig:stab2}
shows the size of the largest absolute value of the
instabilities for each point on the dashed curve of Fig.
\ref{fig:stab}. It illustrates the important fact that the size
of the imaginary parts, which gives the rate of the
instabilities, increases starting from zero, quite rapidly with
$b$, although they remain small as compared with the real parts.

\subsection{Creation of a black/white hole}

The stability diagram of Fig.~\ref{fig:stab} suggests a strategy
for creating a sonic black hole from an initial stable state.
Within the upper subsonic region, the vertical axis
$b=0$\linebreak
\begin{figure}
\begin{center}
\epsfig{file=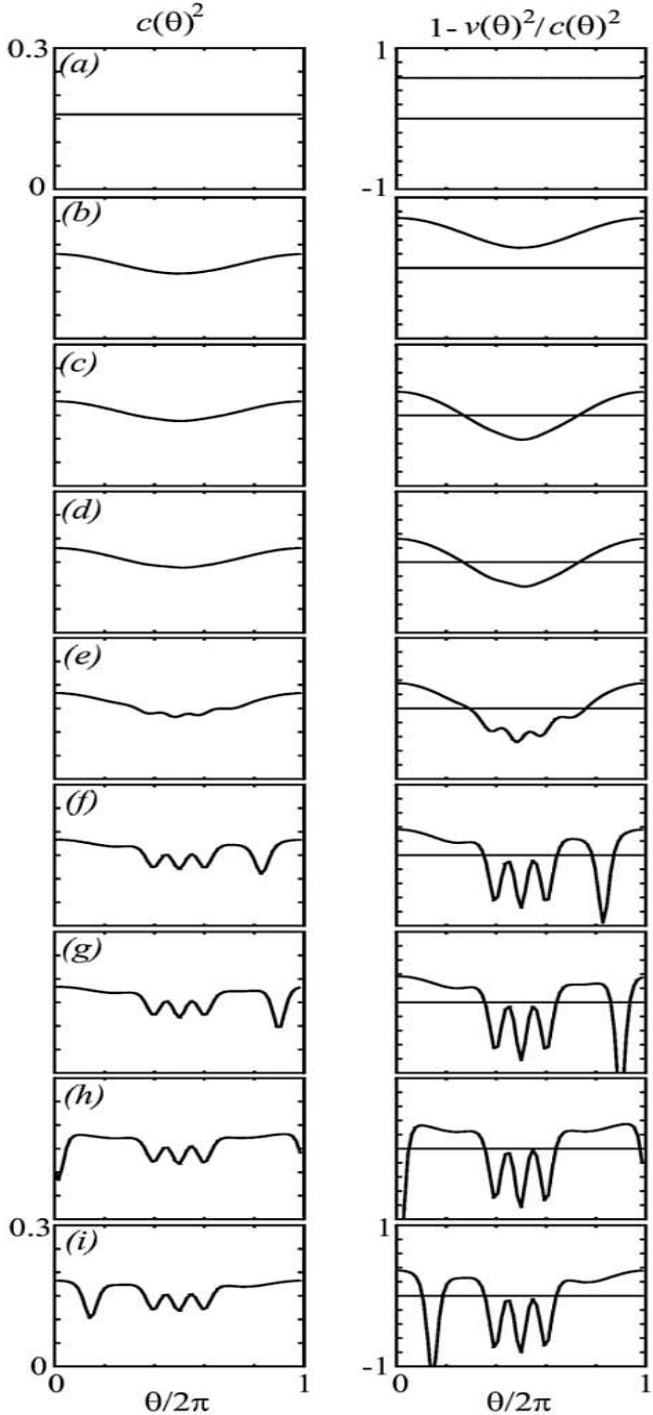,width=\columnwidth,height=.8\textheight}
\end{center}
\caption{Simulation of creation of a stable black/white hole and
subsequent evolution into an unstable region. Figures a--d are
snapshots taken at the initial time (a), at an intermediate time,
still within the subsonic region (b), when the black/white holes
of maximum size is approached (c), and after a long time in that
configuration (d). Then the parameters are changed along the
dashed curve of Fig. \ref{fig:stab} to enter an unstable region
(e) and kept there(f--i). It can be observed that a perturbation
grows at the black hole horizon and travels rightwards until it
enters the white hole horizon.} \label{fig:simul}
\end{figure}\noindent
corresponds to a homogeneous persistent current in a ring, which
can in principle be created using different techniques
\cite{dum98}. Gradually changing ${\cal U}$ and $b$, it is
possible to move from such an initial state to a black/white hole
state, along a path lying almost entirely within the stable
region, and only passing briefly through instabilities where they
are sufficiently small to cause no difficulty.

Indeed, we have simulated this process of adiabatic creation of
a sonic black/white hole by solving numerically (using the split
operator method) the time-dependent Gross-Pitaevskii
equation~(\ref{eq:gp-ring}) that provides the evolution of the
condensate when the parameters of the trapping potential change
so as to move the condensate state along various paths in
parameter space. One of these paths is shown in
Fig.~\ref{fig:stab} (light-grey solid line): we start with a
current at $w=7$, $b=0$, and sufficiently high ${\cal U}$ [Fig.
\ref{fig:simul}(a)]; we then increase $b$ adiabatically keeping
${\cal U}$ fixed until an appropriate value is reached [Fig.
\ref{fig:simul}(b)]; finally, keeping $b$ constant, we decrease
${\cal U}$ adiabatically (which can be physically implemented by
decreasing the radius of the ring trap), until we meet the
dashed contour for black holes of comfortable size [Fig.
\ref{fig:simul}(c)]. Our simulations confirm that the small
instabilities which briefly appear in the process of creation do
not disrupt the adiabatic evolution. The final quantum state of
the condensate, obtained by this procedure, indeed represents a
stable black/white hole. We have further checked the stability
of this final configuration by numerically solving the
Gross-Pitaevskii equation~(\ref{eq:gp-ring}) for very long
periods of time (as compared with any characteristic time scale
of the condensate) and for fixed values of the trap parameters.
This evolution reflects the fact that no imaginary frequencies
are present, as predicted from the mode analysis, and that the
final state is indeed stationary [Fig. \ref{fig:simul}(d)]. Once
the black/white hole has been created, one could further change
the parameters $({\cal U},b)$ so as to move between the unstable
`fingers' into a stable region of higher $b$ (a deeper hole).

\subsection{Quasiparticle pair creation}

Instead of navigating the stable region of parameter space, one
could deliberately enter an unstable region [Fig.
\ref{fig:simul}(e)--\ref{fig:simul}(i)]. In the this case, the
black hole should disappear in an explosion of phonons, which
may be easy to detect experimentally. Such an event might be
related to the evaporation process suggested for real black
holes, in the sense that pairs of quasiparticles are created
near the horizon in both positive and negative energy modes. We
will explain this point briefly; a more detailed exposition is
included in the Appendix.

In the language of second quantization, the perturbation field
operator $ \varphi$
satisfies the linear equation
$$
i\dot{\varphi}=-\frac{1}{2}\varphi''-iv\varphi'+
\left(\frac{1}{2}\frac{c''}{c}-\frac{i}{2}v'+c^2\right)\varphi+c^2\varphi^\dag,
$$
which, taking into account that
$[\varphi(\theta),\varphi^\dag(\theta')]=\delta(\theta-\theta')$,
can be written as
$$
i\dot\varphi=[\varphi,H],
$$
where the Bogoliubov Hamiltonian is
\begin{eqnarray}
H=\int d\theta && \left[
-\frac{1}{2}\varphi^\dag\varphi''-iv\varphi^\dag\varphi'
+\left(\frac{1}{2}\frac{c''}{c}-\frac{i}{2}v'+c^2\right)\varphi^\dag\varphi
\right.\nonumber\\ && \left.
+\frac{1}{2}(\varphi^\dag\varphi^\dag+\varphi\varphi)
\right].
\end{eqnarray}

The Hermiticity of the Bogoliubov linearized Hamiltonian implies
that eigenmodes with complex frequencies
appear always in dual pairs, whose frequencies are complex conjugate. In the
language of second quantization, the linearized Hamiltonian for each such
pair has the form
$$
H=\sum_{n}( \omega A^\dag_{\omega^*,n} A_{\omega,n}
+\omega^* A^\dag_{\omega,n} A_{\omega^*,n}),
$$
and the only nonvanishing commutators among these operators are
$[A_{\omega,n},A^\dag_{\omega^*,n'}]=\delta_{nn'}$. The asterisk
on the subscript is important: the mode with frequency
$\omega^*$ is a different mode from the one with frequency
$\omega$, and $A^\dag_{\omega^*,n}$ is not the Hermitian
conjugate of $A^\dag_{\omega,n}$. It is therefore clear that
none of these operators is actually a harmonic oscillator
creation or annihilation operator in the usual sense. However,
the linear combinations
$$
a_{n}=\frac{1}{\sqrt 2}(A_{\omega,n}+A_{\omega^*,n})\ ,\qquad
b_{n}=\frac{i}{\sqrt 2}(A^\dag_{\omega,n}+A^\dag_{\omega^*,n})
$$
and their Hermitian conjugates are true annhilation and creation operators,
with the standard commutation relations, and in terms of these
the Bogoliubov Hamiltonian  becomes
$$
H = \sum_n\left[{\rm Re}(\omega)(a^\dag_{n} a_{n} -
b^\dag_{n} b_{n}) - {\rm
Im}(\omega)(a^\dag_{n} b^\dag_{n}+
a_{n} b_{n})\right]\;.
$$
This interaction obviously leads to self-amplifying creation of positive and
negative frequency pairs.
Evaporation
through an exponentially self-amplifying instability is not
equivalent, however, to the usual kind of Hawking radiation
\cite{corley+99}; this issue will be discussed in detail
elsewhere.

\section{Sink-generated black holes}

Condensates which develop black hole behaviors by means of flows generated
by laser-driven sinks also  present  regions of stability and instability
in parameter space and, in this sense, their behavior is analogous to that
in a ring-shaped trapping potential. Here we present a simple model that
exhibits the main qualitative features of more general situations and that
can be  studied analytically. Although in this model, we study a
condensate of infinite size, in more realistic models or experiments,
it will suffice to take condensates which are sufficiently large, since
the stability pattern is not significantly affected by the (large but finite)
size of the condensate.

\subsection{The model}

Let us consider a tight cigar-shaped condensate of infinite size
such that the motion in the $(y,z)$ plane is effectively frozen.
In appropriate dimensionless units, the effectively
one-dimensional Gross-Pitaevskii equation thus becomes:
$$
i\partial_\tau\Phi=\left(-\frac{1}{2}\partial_x^2 +{\cal V}_{\rm ext}
+{\cal U}|\Phi|^2\right)\Phi,
$$
with the normalization condition
$$
\lim_{D\to \infty}\frac{1}{2D}\int_{-D}^D dx|\Phi(x)|^2=n.
$$
In this equation, ${\cal V}_{\rm ext}$
is the dimensionless effective potential that results form the dimensional
reduction, which already includes the chemical potential.

In order to obtain a black hole configuration, let us choose the
potential  ${\cal V}_{\rm ext}$ so that it produces
 a profile for the speed of sound $c(x)=\sqrt{{\cal U}\rho(x)}$ of the form
$$
c(x)=
\left\{\begin{array}{ll}
c_0, & |x|<L
\\
c_0 [1+(\sigma-1)x/\epsilon], \quad& L<|x|<L+\epsilon
\\
\sigma c_0, & L+\epsilon<|x|
\end{array}\right. ,
$$
with $\sigma >1$,
and a flow velocity in the inward direction.
The continuity equation then provides the flow velocity profile
$$
v(x)=-\frac{v_0c_0^2}{c(x)^2}\frac{x}{|x|},
$$
where $v_0$ is the absolute
value of the flow velocity in the inner region.

As it stand this model fails to fulfill the continuity equation at $x=0$.
In order to take this into account, we will also introduce a sink of
atoms at $x=0$ that takes atoms out of the condensate (this can be physically
implemented by means of a laser). From the mathematical point of view, it
can be modeled by an additional term in the equation of the form
$-iE\delta(x)$ which indeed induces loss of atoms at $x=0$. Equivalently,
 it can be represented by boundary conditions of the form
\begin{eqnarray}
\Phi(0^+,\tau)-\Phi(0^-,\tau)&=&0,
\nonumber\\
\Phi'(0^+,\tau)-\Phi'(0^-,\tau)&=&-2iE\Phi(0,\tau),
\label{eq:bcphi}
\end{eqnarray}
which determine the flow velocity inside in terms of the
characteristics of the outcoupler laser, namely, $v_0=E$.

Perturbations $\phi$ around this stationary state
$\Phi_s=\sqrt\rho e^{i\int v(x)dx}$, such that
$\Phi=\Phi_s+\phi$ (note that for convenience we have chosen a different
convention as compared with the ring in which
$\Phi=\Phi_s+\varphi e^{i\int v}$) must satisfy the boundary conditions
(\ref{eq:bcphi}) and the equation
$$
i\dot\phi=-\frac{1}{2}\phi''+(c^2-v^2/2+c''/2c)\phi + c^2e^{2i\int^x v}\phi^*
$$
where
$$
\frac{c''}{c}=\frac{\sigma-1}{\epsilon}[\delta(|x|-L)
-\frac{1}{\sigma}\delta(|x|-L-\epsilon) ].
$$
As a further simplifying assumption, we will assume that
$v_0\epsilon \ll 1$ so that
$$
\left|\int_L^{x} dx' v(x')\right|\leq
\int_L^{L+\epsilon} dx\frac{v_0}{[1+(\sigma-1)x/\epsilon]^2}
\leq v_0\epsilon\ll 1.
$$

Let us now expand the perturbation $\phi$ in modes
$$
\phi=\sum_{\omega,k}\left[
A_{\omega,k} u_{\omega,k}(x) e^{-i\omega\tau}+
A_{\omega,k}^* v_{\omega,k}(x)^* e^{i\omega^*\tau}
\right].
$$
Then, the modes  $u_{\omega,k}(x)$ and $v_{\omega,k}(x)$
satisfy, in each region, the Bogoliubov equations
\begin{eqnarray}
\omega u_{\omega,k}&=& -\frac{1}{2}u''_{\omega,k}+(c^2-v^2/2)u_{\omega,k}
+c^2 e^{2i\int^x v}v_{\omega,k},
\nonumber\\
\omega v_{\omega,k}&=& \frac{1}{2}v''_{\omega,k}-(c^2-v^2/2)v_{\omega,k}
-c^2 e^{-2i\int^x v}u_{\omega,k}.
\label{eq:boguv}
\end{eqnarray}

\subsection{Matching conditions}

The intermediate regions $L<|x|<L+\epsilon$, provide the connection between
the perturbation modes in the inner and outer regions. Once these
connection formulas have been established, in the limit of small
$\epsilon$, we will only need to study the inside and outside modes
and their relation through such formulas.  The case of an abrupt horizon, in
which the background condensate velocity is steeply and linearly
ramped within a very short interval, is obviously quite special;
and it does not
particularly
resemble the horizon of a large black hole in Einsteinian gravity.  But the
connection formula that we derive for this case will qualitatively resemble
those that are obtained, with considerably more technical effort, for smoother
horizons\cite{otherpaper}.  And the results we will obtain for the global
Bogoliubov spectrum of the condensate black hole will indeed be representative
of more generic cases.

In the intermediate regions $L<|x|<L+\epsilon$, the factors
$e^{\pm2i\int^x v}$ in the last terms of Eqs. (\ref{eq:boguv})
become $1+{\cal O}(\epsilon)$. Then the solution of these
equations is
\begin{eqnarray}
u_{\omega,k}&=&\alpha_{\omega,k}+\beta_{\omega,k} x/\epsilon
+{\cal O}(\epsilon^2),
\nonumber\\
v_{\omega,k}&=&\gamma_{\omega,k}+\kappa_{\omega,k} x/\epsilon
 +{\cal O}(\epsilon^2)
\label{eq:modesbetween}
\end{eqnarray}
as can be easily seen by defining the variable $q=x/\epsilon$ so that the
equations become
$$
\partial_q^2 u_{\omega,k}=\partial_q^2 v_{\omega,k}={\cal O}(\epsilon^2).
$$

The singular character of $c''/c$ at $|x|=L,\ L+\epsilon$ can be
substituted by matching conditions at $|x|=L,\ L+\epsilon$
which will in turn provide the connection formulas between the modes
outside ($|x|>L$) and the modes inside ($|x|<L$). Furthermore,
the symmetry of the problem allows us to study the region
$x>0$.

These matching conditions are
\begin{eqnarray}
\phi(L^+)-\phi(L^-)&=&0,\nonumber\\
\phi'(L^+)-\phi'(L^-)&=&\frac{\sigma -1}{\epsilon}\phi(L),
\nonumber\\
\phi(L+\epsilon^+)-\phi(L+\epsilon^-)&=&0,
\nonumber\\
\phi'(L+\epsilon^+)-\phi'(L+\epsilon^-)&=&
-\frac{\sigma -1}{\sigma\epsilon}\phi(L+\epsilon).
\nonumber
\end{eqnarray}
These equations together with the form of the modes in the region
$L<x<L+\epsilon$, provide the connection formulas between the inside and
outside modes (from now on we will drop the subindex $\omega$):
\begin{eqnarray}
u_{{\rm in},k}(L)&=& -\epsilon u_{{\rm out},k}'(L)+
\frac{1}{\sigma}u_{{\rm out},k}(L),
\nonumber\\
u_{{\rm in},k}'(L)&=& \sigma u_{{\rm out},k}'(L),
\nonumber
\end{eqnarray}
and likewise for the modes $v_{\rm in,out}$.

\subsection{Dispersion relation and boundary conditions for large $x$}

In each of the regions (inside and outside), we can write
$$
u_{k}(x)=u_{k}e^{i(k-|v|)(x-L)},\quad
v_{k}(x)=v_{k}e^{i(k+|v|)(x-L)}.
$$
Upon substitution of this expansion into the Bogoliubov equations
(\ref{eq:boguv}),
we obtain, for each region,
the following set of algebraic equations:
$$
h_k^- u_k+c^2v_k=0,
\qquad
c^2u_k+h_k^+v_k=0,
$$
where $h_k^\pm=k^2/2+c^2\pm (k|v|+\omega)$. For these equations to have
a solution, the determinant must vanish thus providing
the dispersion relation
\begin{equation}
k^4/4+(c^2-v^2)k^2-2\omega |v|k-\omega^2=0
\label{eq:k4}
\end{equation}
which, for fixed $\omega$, is a fourth-order equation for $k$.
For each of the four solutions $u_{k}$ and $v_{k}$ must be
related by
$$
v_{k}=h_{k}u_{k}\quad\mbox{with}\quad
h_{k}=-\frac{1}{c^2}(k^2/2+c^2-k|v|-\omega).
$$
The constant coefficients $u_{k}$ can be regarded as
normalization constants and will be set to unity. Let us study
the possible solutions to the dispersion relation depending on
whether $\omega$ is real or complex.

\paragraph*{Complex frequencies.---}
In this case all four solutions are pure complex, two of them with
positive imaginary part and two of them with negative one.

In order
to prove this statement, let us first assume that there exists a
real solution $k$ for a complex $\omega=\varpi+i\gamma$. Then
the imaginary part of Eq.~(\ref{eq:k4})  implies that
$\varpi=-k|v|$. Introducing this result into its real part leads
to $\gamma^2=-(k^4/4+c^2k^2)$ which is impossible to  fulfill
because $k$ is real. So the four  solutions are complex.

Because of continuity, all $\omega$ in the upper half complex
$\omega$ plane
 have the same number of solutions with positive imaginary part.
Otherwise, for some $\omega$ there should exist a real solution
that interpolates between positive and negative imaginary part
solutions, but this is not possible, as we have seen.

Now let us concentrate on small frequencies, i.e., on
frequencies around $\omega=0$. For $\omega=0$, we have a double
root at $k=0$. The other two solutions are
$k=\pm2\sqrt{v^2-c^2}$, which are real for $c^2<v^2$ (i.e.,
inside) and pure imaginary for $c^2>v^2$ (i.e., outside). Let us
follow these four solutions when $\omega=i\varepsilon$. The
solutions coming from the double root $k=0$ will now be of the
form
 $k=k_r+i\varepsilon  k_i$. It is easy to see that $k_r=0$ and
$k_i=1/(-|v|\pm c)$. If $c> |v|$, one is positive and one is
negative. If $c< |v|$, both of them are negative. On the other
hand, the solutions $k=\pm2\sqrt{v^2-c^2}$, for $c^2> v^2$, are
already complex conjugate. For $c^2< v^2$, we write
$k=k_r+i\varepsilon k_i$ and introduce it into
Eq.~(\ref{eq:k4}). We then see that at first order in
$\varepsilon$, $k_i=|v|/(v^2-c^2)>0$. Thus, we have seen that
for $\omega=i\varepsilon$, we have two solutions with positive
imaginary part and two with negative one in any case (inside and
outside). But if this is so for $\omega=i\varepsilon$ it must be
true in the whole upper $\omega$ plane and consequently in the
whole complex $\omega$ plane. {\large $\Box$}

In the inside region all possible solutions are in principle allowed but
outside we are only left with the two that have ${\rm Im}(k)>0$,
because the other two grow exponentially.

\paragraph*{Real frequencies.---}
Outside ($c^2>v^2$), there are two real and two complex
conjugate solutions. Of these two complex solutions, only one is
allowed [the one with ${\rm Im}(k)>0$] because the other grows
exponentially. Inside ($c^2<v^2$), for $\omega>\omega_{\rm max}$
there are two real and two complex conjugate solutions; for
$\omega<\omega_{\rm max}$ there are four real solutions; the
value $\omega=\omega_{\rm max}$ is a bifurcating point.

\subsection{Connection formulas for complex frequencies}

Since we are interested in the existence of dynamical
instabilities, we will concentrate in the case in which $\omega$
is complex. Then, as we have seen, the dispersion equation
(\ref{eq:k4}) has four complex solutions for $k$ in each region.
Inside, all four solutions $k_{{\rm in},i},\ i=1\cdots 4$ are in
principle possible but outside those with ${\rm Im}(k_{\rm
out})<0$ will increase exponentially. Therefore, up to
corrections coming from the finite size of the condensate, which
we ignore here, only modes associated with $k_{{\rm
out},\alpha},\ \alpha=1,2$ such that ${\rm Im}(k_{{\rm
out},\alpha})>0$ are allowed. Each mode $u_{{\rm
out},\alpha}(x)= e^{i(k_{{\rm out},\alpha}-v_0/\sigma^2)(x-L)}$
will match a linear combination $u_{{\rm in},\alpha}(x)=\sum_i
F_{\alpha i} u_{{\rm in},i}(x)$ of modes
 $u_{{\rm in},i}(x)$ inside, i.e.,
$$
u_{{\rm in},\alpha}(x) =
\sum_i F_{\alpha i}
e^{i(k_{{\rm in},i}-v_0)(x-L)}
$$
and similarly for
$v_{{\rm out},\alpha}$ and $v_{{\rm in},\alpha}$:
$$
v_{{\rm in},\alpha}(x) =
\sum_i F_{\alpha i} h_{{\rm in},i}
e^{i(k_{{\rm in},i}+v_0)(x-L)}.
$$

After some straightforward calculations, it can be seen that these
 connecting coefficients $F_{\alpha i}$ are given by
$F_{\alpha i}=\sum_j (M^{-1})_{ij}C_{\alpha j}$, where
$$
M=\left(\begin{array}{cccc}
1&1&1&1\\
k_{{\rm in},1}^-&
k_{{\rm in},2}^-&
k_{{\rm in},3}^-&
k_{{\rm in},4}^-\\
h_{{\rm in},1}&
h_{{\rm in},2}&
h_{{\rm in},3}&
h_{{\rm in},4}\\
h_{{\rm in},1}k_{{\rm in},1}^+&
h_{{\rm in},2}k_{{\rm in},2}^+&
h_{{\rm in},3}k_{{\rm in},3}^+&
h_{{\rm in},4}k_{{\rm in},4}^+
\end{array}\right),
$$
$$
C_\alpha=\left(\begin{array}{c}
1/\sigma-i\epsilon k_{{\rm out},\alpha}^-
\\
\sigma k_{{\rm out},\alpha}^-
\\
(1/\sigma-i\epsilon k_{{\rm out},\alpha}^+)
h_{{\rm out},\alpha}
\\
\sigma k_{{\rm out},\alpha}^+
h_{{\rm out},\alpha}\end{array}\right).
$$
In these equations,
$$
k_{{\rm in},i}^\pm=k_{{\rm in},i}\pm v_0\quad \mbox{and}\quad
k_{{\rm out},\alpha}^\pm=k_{{\rm out},\alpha}\pm v_0/\sigma^2.
$$

\subsection{Boundary conditions at the sink. Complex eigenvalues}

We have already found the modes in the inner and outer region as well as
their relation. To determine which (complex-frequency) modes will
be present, there only remains
to impose the boundary conditions dictated by the presence of the
sink at $x=0$.

As we have already mentioned, the symmetry of the system under
reflection ($x\rightarrow -x$) allows us to study only the
region $x>0$, provided that we study the even and odd
perturbations separately. For odd fluctuations [$\phi_{\rm
o}(x,\tau)=-\phi_{\rm o}(-x,\tau)$], the boundary conditions
(\ref{eq:bcphi}) become
$$
\phi_{\rm o}(0,\tau)=0
$$
at all times $\tau$.  This implies that the $u$ and $v$ components of $\phi$
must separately satisfy the boundary condition.  Since we can have any linear
combination of the two solutions that decay outside the horizon, we therefore
have a two-by-two matrix constraint.  The condition that a non-zero solution
exists is that the determinant
$$
{\rm det}\left(\begin{array}{cc}
u_{{\rm in},1}(0) &u_{{\rm in},2}(0)\\
v_{{\rm in},1}(0) & v_{{\rm in},2}(0)
\end{array}\right)=0
$$
and therefore
\begin{equation}
\sum_{ij} F_{1i}F_{2j}(h_{{\rm in},i}-h_{{\rm in},j})
e^{-i(k_{{\rm in},i}+k_{{\rm in},j})L}=0\;.
\label{final1}
\end{equation}

For even fluctuations [$\phi_{\rm e}(x,\tau)=\phi_{\rm e}(-x,\tau)$],
the boundary conditions (\ref{eq:bcphi}) become
$$
\phi_{\rm e}'(0,\tau)+iv_0 \phi_{\rm e}(0,\tau)=0,
$$
which implies that
\begin{equation}
\sum_{ij} F_{1i}F_{2j}(h_{{\rm in},i}-h_{{\rm in},j})
k_{{\rm in},i}k_{{\rm in},j}
e^{-i(k_{{\rm in},i}+k_{{\rm in},j})L}=0.
\label{final2}
\end{equation}

For fixed $L$, ${\cal U}$, $v_0$, and $\sigma$, the quantities
$F$, $h_{\rm in}$, and $k_{\rm in}$ that appear in Eqs.
(\ref{final1}) and (\ref{final2}) are only functions of
$\omega$. Therefore the solutions to these equations are all the
 possible complex eigenfrequencies, which depend on the free
parameters that determine the model, namely, the size $2L$ of
the inner region, the speed of sound inside $c_0$, the relative
change of the speed of sound between the inner and the outer
regions $\sigma$, and the flow velocity inside $v_0$ (related
to the characteristics of the outcoupler laser). In practice,
there are also
other parameters of the condensate such as its size $2D$ (which
 has been made arbitrarily large) and the size of the intermediate
regions $\epsilon$ (which has been made arbitrarily small).

Equations (\ref{final1}) and (\ref{final2}) can be solved
numerically for different values of the parameters $\sigma $,
${\cal U}$, $v_0$, and $L$. The numerical method employed is the
following. The equations above have the form
$$
f(\omega;\sigma, {\cal U}, v_0, L)=0\;,
$$
where $f$ and $\omega$ are both in general complex. We plot
contours of constant absolute value of $f$ in the complex
$\omega$ plane; where $|f|$ approaches zero, we have an
eigenfrequency.

The distribution of complex solutions in the complex $\omega$
plane depends on the size of the inner region $L$, for given
$\sigma$, ${\cal U}$, and $v_0$. Direct inspection of the
numerical results shows that the number of instabilities
increases by one when the black-hole size $L$ is increased by
$\pi/k_0$ where $k_0=\sqrt{v_0^2-c_0^2}$. More explicitly, for
$L$ smaller than $\pi/k_0-\delta$ ($\delta$ being much smaller
than $\pi/k_0$) there are no complex eigenfrequencies; for
$(L+\delta)k_0/\pi\in[n,n+1]$ with $n=1,2\ldots$, we have $n$
complex solutions except for $L=(n+1/2)\pi/k_0$, where we find
$n-1$ complex solutions instead of $n$ [i.e, there is one mode
for which ${\rm Im}(\omega)=0$ within numerical resolution].
This can be easily interpreted qualitatively since the unstable
modes are basically the bound states in the black hole, and the
highest wave number $k$ on the positive norm upper branch, for
the barely bound state with $\omega\to0^-$, is exactly $k_0$. So
the threshold is simply when the well becomes big enough to have
a bound state; the small $\delta$ displacement comes in because
the horizon is not exactly a hard wall; and similarly for the
extra bound state every $\pi/k_0$. Thus stability can only be
achieved for small sizes of the inner region, $L\lesssim
\pi/k_0$. As we discussed in Sec. II, the wave length $2\pi/k$
of the perturbations must be smaller than this size, which
implies $k> 2\pi/L\gtrsim 2 k_0$. However, for these
perturbations the hydrodynamic approximation, which requires
$k\lesssim 2k_0$, is not valid. Therefore there are no stable
black hole configurations in a strict sense. The sizes of the
imaginary parts of the complex solutions decrease as the size
$L$ of the interior of the black hole increases. Thus although a
larger hole has more unstable modes, it is actually less
unstable (and might even became quasistable in the sense that
its instability time scale would be longer than the experimental
duration).

\section{Conclusions}

We have seen that dilute Bose-Einstein condensates admit, under
appropriate conditions, configurations that closely ressemble
gravitational black holes. We have analyzed in detail the case
of a condensate in a ring trap, and proposed a realistic scheme
for adiabatically creating stable sonic black/white holes and we
have seen that there exist stable and unstable black hole
configurations. We have also studied a model for a
sink-generated sonic black hole in an infinite one-dimensional
condensate. The dynamical instabilities can be interpreted as
coming from quasiparticle pair creation, as in the well-known
suggested mechanism for black hole evaporation. Generalizations
to spherical or quasi-two-dimensional traps, with flows
generated by laser-driven atom sinks, should also be possible,
and should behave similarly. While our analysis has been limited
to Bogoliubov theory, the further theoretical problems of back
reaction and other corrections to simple mean field theory
should be more tractable for condensates than for other systems
analogous to black holes. And we expect that experiments along
the lines we have proposed, including both creation and
evaporation of sonic black holes, can be performed with
state-of-the-art or planned technology.

\acknowledgments

We thank the Austrian Science Foundation and the European Union
TMR networks ERBFMRX--CT96--0002 and ERB--FMRX--CT96--0087.
J.R.A. is grateful to Ted Jabobson for useful discussions.

\appendix

\section*{Complex frequencies. Redundancy and normalization}
In this Appendix, we will analyze the issue of the redundancy
and normalization of the Bogoliubov modes in the presence of
complex frequencies from a general point of view.  Dynamical
instabilities in quantum field theory, and the quantization of
dynamically unstable modes, do not seem to be widely understood:
it is for instance common to read axiomatic statements that one
must only quantize positive norm modes, even though this
implicitly neglects dynamical instabilities, and does not
follow in general from the fundamental commutation relations.
But some explicit treatments of quantum instabilities have been
available in the literature for some time\cite{DIQFT}; here we
review this subject in the specific Bogoliubov context.

We will begin by writing the Bogoliubov equations in their most
usual form
\begin{equation}\label{Bog}
{\hbar\omega_j}\left(\matrix{u_j\cr v_j}\right) =
\left(\matrix{
h_0({\bf x})
& c({\bf x})^2 e^{2i\vartheta({\bf x})}\cr
- c({\bf x})^2 e^{2i\vartheta({\bf x})}
& -h_0({\bf x})}
\right)\;\left(\matrix{u_j\cr v_j}\right)
\;,
\end{equation}
where $h_0({\bf x})=-{\hbar^2/2m}\nabla^2 +V_{\rm ext}({\bf x})
+ 2m c({\bf x})^2- \mu$. In terms of these modes, the atomic
second quantized field operator has the well-known form
$\hat\Psi({\bf x},t)=\Psi_s({\bf x},t)+\hat \psi({\bf x},t)$
with
\begin{equation}
\hat \psi({\bf x},t)=\sum_j\left[
\hat a_j u_j({\bf x})e^{-i \omega_j t}+
\hat a_j^\dag v_j({\bf x})^*e^{i \omega_j^* t}\right].
\label{delpsiexp}
\end{equation}

If there is a solution $(u_j,v_j)$ to Eq. (\ref{Bog}) with mode frequency
$\omega_j$, then straightforward substitution shows that $(u_{j'},v_{j'}) =
(v_j^*,u_j^*)$ must be a solution with frequency $\omega_{j'}=-\omega_j^*$.
If we examine the contributions of these two solutions,
however, we find that together they yield but a single term, of the form
$(\hat{a}_j + \hat{a}_{j'}^\dagger) e^{-i\omega t} u_j({\bf x}) +
(\hat{a}_j^\dagger + \hat{a}_{j'}) e^{i\omega^* t} v_j^*({\bf x})$.  It is
thus a quite trivial fact that the two modes $j$
and $j'$ are redundant.  We are free to simplify our notation by
redefining $\hat{a}_j + \hat{a}_{j'}^\dagger \to \hat{a}_j$, and eliminating
mode $j'$ (leaving it out of the sum over frequencies).  Alternatively we
could of course eliminate $j$ and keep $j'$.  Which of these two notational
conventions we should take is best determined by the commutator $[\hat{a}_j +
\hat{a}_{j'}^\dagger, \hat{a}_j^\dagger + \hat{a}_{j'}]$, which will tell us
whether the coefficient of $u_j = v_{j'}^*$ is properly an
annihilation operator or a creation operator.

Since the only commutation relations that we are given are those of
$\hat{\psi}$ and $\hat\psi^\dagger$, we must derive the
orthogonality relation for solutions of Eq. (\ref{Bog}), and use it to invert
Eq. (\ref{delpsiexp}).  We can use Eq. (\ref{Bog}) to show that
\begin{eqnarray}\label{orthog1}
(\omega_j + \omega_k) M_{jk}&\equiv&\int\!d^3x\, (u_j v_k - v_j
u_k) =0,
\nonumber\\
(\omega_j - \omega_k^*)
N_{jk}&\equiv&(\omega_j - \omega_k^*)\int\!d^3x\,(u_j u_k^* -
v_j v_k^*) = 0\;,
\end{eqnarray}
where in the case of infinite volume the right hand sides are zero in the
distributional sense, being infinitely rapidly oscillating boundary terms.
This obviously implies that $M_{jk}$ vanishes unless $\omega_k=-\omega_j$,
and $N_{jk}$ vanishes unless $\omega_k=\omega_j^*$.
One can then show that it is always possible to take linear
combinations among degenerate modes, and to eliminate redundant modes as just
discussed, in such a way as to make $M_{jk}$ always vanish, and $N_{jk} =
\delta_{k\bar{\jmath}}$, where for every $j$ there is a single dual mode
$\bar{\jmath}$, with $\omega_{\bar\jmath}=\omega_j^*$.  In the case of real
$\omega_j$, but only then, we have $\bar{\jmath}=j$.  In general, though,
duality
is reciprocal (the dual mode of $\bar{\jmath}$ is always $j$).

The result is that we can now insert Eq. (\ref{delpsiexp}) into
the second quantized Hamiltonian, with the $T$-matrix
approximation for the interparticle interaction, to obtain the
linearized Bogoliubov Hamiltonian for the perturbations
\begin{equation}\label{Hlin}
\hat{H} = \hbar\sum_j \omega_j \hat{a}^\dagger_{\bar{\jmath}}\hat{a}_j\;.
\end{equation}
Since the sum over all modes $j$ also includes the dual to every mode with
complex $\omega_j$, $\hat{H}$ is manifestly Hermitian even though
$\omega_j$ need not be real.  We can also invert Eq. (\ref{delpsiexp}) to learn
that
\begin{equation}\label{delpsiinv}
\hat{a}_j = \int\!d^3x\,[u^*_{\bar{\jmath}} \hat{\delta\psi} +
v_{\bar{\jmath}}\hat{\delta\psi}^\dagger]\;,
\end{equation}
which with Eq. (\ref{orthog1}) implies the commutation relations
\begin{equation}\label{CCR}
[\hat{a}_j,\hat{a}^\dagger_k] = \delta_{k{\bar{\jmath}}},\qquad
[\hat{a}_j,\hat{a}_k] = 0 \;.
\end{equation}

For all $j$ with real $\omega_j$, Eq. (\ref{CCR}) are merely the
standard canonical commutation relations; and our normalization
conventions $M_{jk}=0$ and $N_{jk}=\delta_{jk}$ are likewise the
ones most often presented. In the case of complex $\omega_j$
where $\bar{\jmath}\not= j$, however, Eq. (\ref{CCR}) implies
that the {\it canonical} conjugate of $\hat{a}_j$ is
$\hat{a}^\dagger_{\bar{\jmath}}$, and this is no longer the same
as the {\it Hermitian} conjugate $\hat{a}_j^\dagger$. In fact
for complex $\omega_j$ we have $[\hat{a}_j^\dagger,\hat{a}_j]
=0$; this already follows from the second line of Eq.
(\ref{orthog1}), which implies that the norm $N_{jj}$ of any
mode with complex $\omega_j$ is zero. But if $\hat{a}_j$ and
$\hat{a}_j^\dagger$ commute, then it is clear that neither
$\hat{a}_j$ nor $\hat{a}_{\bar{\jmath}}$ is really a harmonic
oscillator annihilation operator in the usual sense, nor are
$\hat{a}_j^\dagger$ or $\hat{a}_{\bar{\jmath}}^\dagger$ proper
creation operators. The commutation relations (\ref{CCR}) are
validly derived from the fundamental commutation relations for
$\hat{\psi}$ and $\hat{\psi}^\dagger$; but they do not imply,
for instance, that either $\hat{a}^\dagger_j\hat{a}_j$ or
$\hat{a}_{\bar{\jmath}}^\dagger\hat{a}_j$ has the discrete,
equally spaced spectrum, bounded from below, that one expects of
a quasiparticle number operator.

To understand the dual pairs of modes with complex frequencies, we can define
the ordinary annihilation operators
\begin{equation}\label{bbbar}
\hat{b}_j ={1\over\sqrt{2}}(\hat{a}_j+\hat{a}_{\bar{\jmath}}),\qquad
\hat{b}_{\bar{\jmath}} =
{i\over\sqrt{2}}(\hat{a}_j^\dagger-\hat{a}_{\bar{\jmath}}^\dagger)
\end{equation}
and their Hermitian conjugates, among which the only
nonvanishing commutators are the ordinary
\begin{equation}\label{CCR2}
[\hat{b}_j,\hat{b}_j^\dagger] =
[\hat{b}_{\bar{\jmath}},\hat{b}_{\bar{\jmath}}^\dagger]=1\;.
\end{equation}
In terms of these operators, which are harmonic oscillator annihilation and
creation operators with all the familiar properties of such, the
$j,\bar{\jmath}$
subsector of the Bogoliubov Hamiltonian $\hat{H}$ appears as
\begin{equation}\label{Hjjb}
\hat{H}_{j,\bar{\jmath}} = {\rm Re}(\omega_j) [\hat{b}^\dagger_j\hat{b}_j -
\hat{b}^\dagger_{\bar{\jmath}}\hat{b}_{\bar{\jmath}}] - {\rm Im}(\omega_j)
[\hat{b}^\dagger\hat{b}_{\bar{\jmath}}^\dagger +
\hat{b}_j\hat{b}_{\bar{\jmath}}]\;.
\end{equation}
  Note that Eq. (\ref{Hjjb}) is only the
simplest form in which one may write the $j,\bar{\jmath}$ sector
of the Hamiltonian: by introducing appropriate factors of
$e^{\pm i\alpha/2}/\cos\alpha$ into Eq. (\ref{bbbar}), for any
$\alpha$, we can make ${\rm Im}(\omega_j)\to {\rm
Im}(\omega_j)/\cos\alpha$ and add a term ${\rm
Im}(\omega_j)\tan\alpha (\hat{b}^\dagger_j\hat{b}_j +
\hat{b}^\dagger_{\bar{\jmath}}\hat{b}_{\bar{\jmath}})$.

We can now examine the spectrum of $\hat{H}_{j,\bar{\jmath}}$ by considering it
in
the basis of eigenstates of $\hat{n} = \hat{b}^\dagger_j\hat{b}_j +
\hat{b}^\dagger_{\bar{\jmath}}\hat{b}_{\bar{\jmath}}$ and $\hat{\Delta} =
\hat{b}^\dagger_j\hat{b}_j -
\hat{b}^\dagger_{\bar{\jmath}}\hat{b}_{\bar{\jmath}}$.  In
fact $\hat{\Delta}$ commutes with $\hat{H}_{j,\bar{\jmath}}$, so
defining
\begin{eqnarray}
|E_\Delta\rangle & = & \sum_{n=0}^\infty c_n|n+\Delta\rangle|n\rangle\quad
\Delta\geq 0\;,
\nonumber\\
|E_\Delta\rangle & = & \sum_{n=0}^\infty c_n
|n\rangle |n+\Delta\rangle\quad \Delta\leq 0\;,
\label{PhiDel}
\end{eqnarray}
where $\hat{b}^\dagger_j\hat{b}_j |m\rangle |n\rangle = m |m\rangle
|n\rangle$ and $\hat{b}^\dagger_{\bar{\jmath}}\hat{b}_{\bar{\jmath}}|m\rangle
|n\rangle
= n |m\rangle |n\rangle$, we find $\hat{H}_{j,\bar{\jmath}}|E_\Delta\rangle =
[\hbar\Delta {\rm Re}(\omega_j) + E_\Delta]|E_\Delta\rangle$.  And we have
the recursion relation
\begin{eqnarray}
E_\Delta c_n = {\rm Im}(\omega_j)[&&\sqrt{n(n+\Delta)} c_{n-1}
\nonumber\\
&&+
\sqrt{(n+1)(n+\Delta+1)}c_{n+1}]\;.
\label{recurse}
\end{eqnarray}
As $n\to\infty$, we have $c_{n+1}\to - c_{n-1}$, and so $\sum_n |c_n|^2$ does
not converge: none of the eigenstates of $\hat{H}_{j,\bar{\jmath}}$ is
normalizable.  One can however obtain delta-function normalization for a
continuous spectrum of real $E_\Delta$, bounded neither above nor below.

That the Hamiltonian $\hat{H}$ is unbounded from below does not
indicate anything unphysical about our model: we have simply linearized about
an unstable excited state of the nonlinear full Hamiltonian, which is
bounded from below.  Real negative frequencies $\omega_j$, where our
convention $N_{jj}=1$ has been imposed, indicate energetic instabilities,
whereby the system will decay in the presence of dissipation.  Complex
$\omega_j$, on the other hand, indicate dynamical instabilities.
Classically, a dynamically unstable system will exponentially diverge from
the initial stationary state if is perturbed, even without dissipation.
Quantum mechanically, we have just seen that a dynamically unstable system
has no normalizable stationary states.  If an initially stable system is
driven into a state which is stationary but dynamically unstable at the
classical (mean field) level, the initial state will have had finite Hilbert
space norm, and hence under unitary evolution the final state will have the
same norm.  Thus it will not be a stationary state; one may say that quantum
fluctuations will always trigger the dynamical instability.  For a
logarithmically long period of time, however, the linearized theory will
still remain valid.  In this sense, our linearized description of quantum
dynamical instabilities is sound.


\begin{references}

\bibitem{BECexp95} M. H. Anderson, J. R. Ensher, M. R. Matthews,
C. E. Wieman, and E. A. Cornell, {Science} {\bfseries 269}, 198
(1995); K. B. Davis {\it et al.}, {Phys. Rev. Lett.} {\bfseries
75}, 3969 (1995).

\bibitem{dalfovo+99}
See, e.g., F. Dalfovo, S. Giorgini, L. P. Pitaevskii, and S.
Stringari, {Rev. Mod. Phys.} {\bfseries 71}, 463
(1999).

\bibitem{volovik99} For a review, see, e.g.,
G. E. Volovik, in {\it Topological Defects and the
Non­-Equilibrium Dynamics of Symmetry Breaking Phase
Transitions}, edited by Y.M. Bunkov and H. Godfrin (Kluwer
Academic Publishers, 2000) p. 353; V. B. Eltsov, M. Krusius, and
G. E. Volovik, cond-mat/9809125 (unpublished).


\bibitem{unruh81}
W.G. Unruh, {Phys. Rev. Lett.} {\bfseries 46}, 1351
(1981).

\bibitem{unruh95}
W. G. Unruh, {Phys. Rev.} D {\bfseries 51}, 2827 (1995).

\bibitem{visser} M. Visser, gr-qc/9311028 (unpublished);
S. Liberati, S. Sonego, and M. Visser, Class. Quantum Grav.
{\bfseries 17}, 2903 (2000).

\bibitem{visser98}
M. Visser, {Phys. Rev. Lett.} {\bfseries 80}, 3436
(1998); {Class. Quant. Grav.} {\bfseries 15}, 1767
(1998).

\bibitem{misner+73} C. W. Misner, K. S. Thorne, and J. A. Wheeler,
{\itshape Gravitation} (Freeman, San Francisco, 1973).



\bibitem{hawking74} S. W. Hawking, {Nature (London)} {\bfseries 248},
30 (1974); {Commun. Math. Phys.} {\bfseries 43}, 199
(1975).



\bibitem{jacobson91}
T. Jacobson, {Phys. Rev.} D {\bfseries 44}, 1731 (1991).

\bibitem{corley+99}
S. Corley and T. Jacobson, {Phys. Rev.} D {\bfseries 59}, 4011
(1999); S. Corley, {\it ibid.} {\bfseries 57}, 6280 (1998).



\bibitem{jacobson00} For a review, see, e.g.,
T. Jacobson, Prog. Theor. Phys. Suppl. {\bfseries 136}, 1
(1999).


\bibitem{ruutu+96}
V. M. H. Ruutu {\it et al.}, {Nature (London)} {\bfseries 382},
334 (1996);
T. A. Jacobson and G. E. Volovik, {Phys. Rev.} D {\bfseries 58},
4021 (1998); G. E. Volovik, {Pis'ma Zh. Eksp. Teor. Fiz.}
{\bfseries 69}, 662 (1999); JETP Lett. {\bfseries 69}, 705
(1999).


\bibitem{reznik97}
B. Reznik, gr-qc/9703076 (unpublished).

\bibitem{leonhardt+00}
U.~Leonhardt and P.~Piwnicki, {Phys. Rev. Lett.}
{\bfseries 84}, 822 (2000); {Phys. Rev.} A
{\bfseries 60}, 4301 (1999).


\bibitem{cornell99}
M. R. Matthews {\it et al.}, {Phys.
  Rev. Lett.} {\bfseries  83},  2498  (1999);
L. Denget {\it et al.} {Nature (London)} {\bfseries 398}, 218
(1999);
S. Burger {\it et al.}, {Phys. Rev. Lett.} {\bfseries 83}, 5198
(1999).

\bibitem{bhbec} L. J.~Garay, J. R.~Anglin, J. I.~Cirac,
and P.~Zoller, Phys. Rev. Lett. {\bfseries 85}, 4643 (2000).

\bibitem{andrews97}
M. R. Andrews {\it et al.}, {Science} {\bfseries 275}, 637
(1997); I. Bloch, T. W. H\"ansch, and T. Esslinger, {Phys. Rev.
Lett.} {\bfseries 82}, 3008 (1999); E. W. Hagley {\it et al.},
{Science} {\bfseries 283}, 1706 (1999).


\bibitem{winiecki} See, e.g., T. Winiecki {\it et al.},
Europhys. Lett. {\bfseries 48}, 475 (1999).


\bibitem{khawaja} See, e.g., U. Al Khawaja {\it et al.},
Phys. Rev. A {\bfseries 60}, 1507 (1999); D. L. Feder {\it et
al.}, {\it ibid.} {\bfseries 61}, 011601 (2000).

\bibitem{svistunov} Yu. Kagan, N. Prokof'ev, and B. V.
Svistunov, Phys. Rev. A {\bfseries 61}, 045601 (2000).

\bibitem{otherpaper}  L. J.~Garay, J. R.~Anglin, J. I.~Cirac,
and P.~Zoller, (unpublished).


\bibitem{shlyapnikov}
P. O. Fedichev and G. V. Shlyapnikov, {Phys. Rev.} A
{\bfseries 60}, R1779 (1999).

\bibitem{dum98} R. Dum, J. I. Cirac, M. Lewenstein, and P. Zoller,
{Phys. Rev. Lett.} {\bfseries 80}, 2972 (1998);
J. Williams and M. Holland, {Nature (London)} {\bfseries 401},
568 (1999).


\bibitem{DIQFT} See Gungwon Kang, hep-th/9603166 (unpublished)
 for a concise
pedagogical illustration, and references therein, especially S.
A.~Fulling, {\it Aspects of Quantum Field Theory in Curved
Spacetime} (Cambridge University Press, Cambridge, 1989);
B.~Schroer and J. A.~Swieca, Phys. Rev. D {\bf 2}, 2938 (1970).


\end{references}
\end{document}